\documentclass[prd,twocolumn,nopacs,floatfix,amsmath,nofootinbib,amssymb,preprintnumbers]{revtex4-2}
\usepackage{graphicx,subfigure,color,dcolumn,booktabs}
\usepackage{txfonts}
\usepackage{slashed}
\usepackage{epstopdf}
\usepackage{graphicx,color,dcolumn,booktabs}
\usepackage[colorlinks,
            citecolor=blue,
            anchorcolor=red,
            menucolor=red,
            linkcolor=red,
            filecolor=red,
            runcolor=red,
            urlcolor=blue,
            frenchlinks=red]{hyperref}

\graphicspath{{Figures/}}

\begin{document}

\title{Toy model to understand oscillatory behavior in timelike nucleon form factors}
\author{Ri-Qing Qian$^{1,2,3,4}$}\email{qianrq21@lzu.edu.cn}
\author{Zhan-Wei Liu$^{1,2,3,4}$}
\email{liuzhanwei@lzu.edu.cn}
\author{Xu Cao$^{2,3,5,6}$}\email{caoxu@impcas.ac.cn}
\author{Xiang Liu$^{1,2,3,4}$}\email{xiangliu@lzu.edu.cn}
\affiliation{
$^1$School of Physical Science and Technology, Lanzhou University, Lanzhou 730000, China\\
$^2$Research Center for Hadron and CSR Physics, Lanzhou University and Institute of Modern Physics of CAS, Lanzhou 730000, China\\
$^3$Lanzhou Center for Theoretical Physics, Key Laboratory of Theoretical Physics of Gansu Province and Frontiers Science Center for Rare Isotopes, Lanzhou University, Lanzhou 730000, China\\
$^4$Key Laboratory of Quantum Theory and Applications of MoE, Lanzhou University,
Lanzhou 730000, China\\
$^5$Institute of Modern Physics, Chinese Academy of Sciences, Lanzhou 730000, China\\
$^6$University of Chinese Academy of Sciences, Beijing 100049, China}

\begin{abstract}
To understand the oscillatory behavior exhibited in the timelike electromagnetic form factors of nucleons, we propose a toy model based on the Jost function of the $N\bar N$ pair into the timelike form factors with the help of the distorted-wave Born approximation. By constructing a simple square-well potential reflecting the final-state interaction of $N\bar N$, we naturally represent the damped oscillatory phenomenon in the timelike electromagnetic form factors of nucleons. Especially, our study reveals that the ``period" of the oscillation is approximately determined by the Yukawa interaction range $1/m_\pi$.
Other possible potentials are also discussed.
The threshold enhancements of the cross sections for $e^+e^-\to n\bar n$, $\Lambda\bar\Lambda$, and $\Lambda_c\bar\Lambda_c$ can also be understand within this scenario.
\end{abstract}

\maketitle

\noindent{\it Introduction.}---Nucleon electromagnetic form factors (EMFFs) are of fundamental interest with the aim to understand the internal structure of this subatomic entity (see Refs.~\cite{Denig:2012by,Pacetti:2014jai,Punjabi:2015bba} for recent reviews). They describe the couplings of virtual photon with nucleon electromagnetic current, which is closely related to the dynamical properties of internal quarks and gluons. At small momentum, the EMFFs describe the gross electromagnetic properties of nucleon. At large momentum, they are related to the quark dynamics inside nucleon.

As a function of the 4-momentum squared $q^2$ of virtue photon, the EMFFs of nucleon have two physically separated region.
The EMFFs in spacelike region ($q^2<0$) have been obtained from precise electron-nucleon scattering data with large statistics~\cite{Puckett:2011xg,Puckett:2017flj} and have been widely investigated. However, it is only in recent years that precise data on timelike ($q^2>2m_N$) EMFFs have become available via the measurements of the $e^+e^-$ annihilation process~\cite{BaBar:2013ves,BaBar:2013ukx,BESIII:2019hdp,SND:2022wdb}.
These precise measurements provide a good opportunity to study the EMFFs in the two regions simultaneously. Recently, a combined analysis of the EMFFs in both regions was performed using dispersion theory to extract the electromagnetic radius of the proton~\cite{Lin:2021xrc}.

The precise measurements on the timelike EMFFs of nucleon also brought some puzzles. As first pointed out in Ref.~\cite{Bianconi:2015owa}, the timelike EMFFs of proton show an unexpected oscillation behavior in the near-threshold region. This oscillation feature was later confirmed by the BESIII measurements~\cite{BESIII:2019hdp}, and new data show that a similar oscillation phenomenon exists in the neutron sector~\cite{BESIII:2021tbq,SND:2022wdb}. The effective form factors $G_{\mathrm{eff}}$ of the nucleons were found to be well divided into two parts. The main part $G^0$ can be obtained with a perturbative QCD parametrization and describes the main decreasing behavior of the form factor very well, while the remaining part $G^{osc}$ exhibits a damped oscillation with regularly spaced maxima and minima over the sub-GeV scale~\cite{Bianconi:2015owa}.

The oscillation phenomenon of the EMFF of nucleons has been discussed in different approaches~\cite{Cao:2021asd,Lorenz:2015pba,Bianconi:2015vva,Yang:2022qoy}. The crests and troughs of the oscillation pattern can be explained by introducing some meson resonances~\cite{Cao:2021asd,Lorenz:2015pba}.
The fit in Ref.~\cite{Cao:2021asd} shows that the $\Gamma_{N\bar{N}}$ of the hypothetical mesons will be large compared to $J/\psi$ and $\psi^\prime$, this may cause potential problems for this explanation and needs further research.
A good separation of the damped oscillatory part from a smooth part indicates two different intrinsic mechanisms. A natural speculation is that these two correspond to the bare formation of $N\bar{N}$ and the rescattering of the $N\bar{N}$ pair with the final-state interaction (FSI)~\cite{Bianconi:2015owa}. A qualitative analysis shows that the oscillation can be  produced by an optical potential of a special form~\cite{Bianconi:2015vva}.
A complete and plausible understanding of the emergence of the oscillation behavior is still lacking. A more natural explanation for the observed good periodicity is needed .

In the present work, we propose a toy model, which focus on how the interaction between $N\bar{N}$ pair leads to the oscillation feature. The effect of the interaction between $N\bar{N}$ on EMFFs is closely related to the zero-point wave function of $N\bar{N}$ due to the small interaction range for the formation process compared to that for rescattering. Our study shows that the oscillationlike behavior arises naturally from the rescattering of $N\bar N$ with a well ranged potential of square-well shape, and the ``period" of the oscillation is approximately determined by the Yukawa interaction range $1/m_\pi$. As a by-product, our approach can also interpret the threshold enhancement of timelike form factors.

\noindent{\it Toy model}.---We note that the interaction between $N\bar{N}$ in the $e^+e^-\to N\bar{N}$ process should be different from those obtained directly from $N\bar{N}\to N\bar{N}$ experiments. One obvious observation is that the $N\bar{N}\to N\bar{N}$ proceed with partial or no annihilation of a quark-antiquark pair, while the $N \bar{N}\to e^+e^-$ process requires the annihilation of three quark-antiquark pairs, hence a perfect overlap of wave functions. For this reason, we might expect a separation between the short-range formation and the long-range interaction between $N\bar{N}$.
This separation allows us to study the FSI modification of the production process $e^+e^-\to N\bar{N}$. It is more intuitive to understand this effect for the time reversal process $N\bar{N}\to e^+e^-$, since the attractive interaction between $N\bar N$ will increase the probability that they meet each other and thus increase the cross section, while the repulsive interaction will suppress it.

The formation of the $N\bar{N}$ pair by the virtual photon is of short-range nature, and a naive estimate is $r_0\sim 1/(2m_N)$. While the interaction range associated to $N \bar{N}$ rescattering is approximately given by $a\sim 1/m_\pi$. An estimate of $a/r_0\approx 14$ indicates an explicit separation of formation and rescattering processes. In this sense, we can write the total cross section as:
\begin{equation}\label{eq:cross_section}
  \sigma = \frac{1}{|{\mathcal J}(p)|^2}\sigma_0\,.
\end{equation}
From the distorted-wave Born approximation, the enhancement or suppression factor
${\mathcal J}(p)$ is the Jost function of the final states $N\bar N$ with the pure FSI $V$~\cite{Taylor:1972}, and it is closely related to the zero-point wave function.

The radial wave function $\psi_{\ell,p}(r)$ of $N\bar N$ can be given by the radial Schr\"{o}dinger equation for each $\ell$ partial wave
\begin{equation}\label{eq:shrodinger}
  \left( \frac{d^2}{dr^2}-\frac{\ell(\ell+1)}{r^2}-2\mu V +p^2 \right)\psi_{\ell,p}(r) = 0\,,
\end{equation}
where $\mu$ is the reduced mass and $p$ is the center-of-mass momentum of $N\bar{N}$ system. In S-wave dominant processes,
\begin{equation}\label{eq:jost}
  \mathcal J(p) \approx \mathcal J_{\ell=0}(p)=\lim_{r\to 0} \frac{j_{0}(pr)}{\psi_{0,p}(r)} \,.
\end{equation}
The regular spherical Bessel function $j_0(pr)=\sin(pr)$ is the free radial solution of Eq.~\eqref{eq:shrodinger}. From Eq.~(\ref{eq:jost}), this enhancement/suppression factor $|1/\mathcal J|^2$ has a good probability interpretation, i.e., the ratio of the probability density for finding the scattering state $N\bar N$ near the origin with FSI to that without FSI.

The interaction range of the annihilation and recreation processes of $N\bar N$ is close to $r_0$, and they should be strongly involved in the initial formation of the $N\bar{N}$ pair. Therefore, we also include their contribution in $\sigma_0$, i.e., the FSI $V$ does not contain the annihilation potential and therefore we neglect its imaginary part in this work.

The Coulomb interaction between the $N\bar N$ pair is weaker and has a much larger interaction range compared to the strong interaction for $N\bar N$ rescattering, and this leads to the additional well-known Sommerfeld factor $C=|1/\mathcal S|^2$ with ~\cite{Arbuzov:2011ff}
\begin{equation}\label{eq:sommerfeld}
  \mathcal S = \left(\frac{y}{1-e^{-y}}\right)^{-1/2}\,,\quad y=\frac{\pi\alpha\sqrt{1-\beta^2}}{\beta}.
\end{equation}
Here, $\alpha$ is the electromagnetic fine-structure constant, $\beta = \sqrt{1-4m_N^2/s}$ is the center-of-mass velocity of nucleon, and $s$ is the Mandelstam variable. Actually, we can repeat this Coulomb enhancement factor very easily in our approach. Substituting the Coulomb potential in Eq.~\eqref{eq:shrodinger} for $V$, Eq. (\ref{eq:jost}) is reduced to $\mathcal S$ by using the nonrelativistic approximation $\sqrt{1-\beta^2}\approx 1$ in the near-threshold region.

First, we consider the simple case of a rectangular potential well:
\begin{equation}
  V(r) = \begin{cases}
        -V_a & \mbox{for}\quad 0\leqslant r < a \\
        0 & \mbox{for}\quad r \geqslant a
      \end{cases}
      \,.
\end{equation}
We can easily adjust the interaction range $a$ and the potential depth $V_a$ to glimpse some features of a somewhat complicated potential. The Schr\"{o}dinger equation with this potential has an analytical solution. The general solution for $\ell = 0$ is
\begin{equation}
    \psi_{0,p}(r) =
    \begin{cases}\displaystyle
     \frac{ e^{i\delta_0}\, \mathrm{sin}(p_{in} r)}{\sqrt{\mathrm{sin}^2(p_{in}a)+\frac{p^2}{p_{in}^2}\mathrm{cos}^2(p_{in}a)}} &  \mbox{for}\quad 0\leqslant r < a \\
     e^{i\delta_0}\, \mathrm{sin}(p r+\delta_0) & \mbox{for}\quad r \geqslant a
  \end{cases}\,,
\end{equation}
where $\delta_0$ is the $S$-wave phase shift, and
\begin{equation}
p_{in} = \sqrt{p^2+2\mu V_a}.
\end{equation}
We have the enhancement factor
\begin{equation}\label{eq:enhancement}
  |\mathcal J(p)|  =\sqrt{\frac{p^2}{p_{in}^2}\mathrm{sin}^2(p_{in}a)+\mathrm{cos}^2(p_{in}a)}\,.
\end{equation}
The attractive potential gives an enhancement factor as one can easily check that $|1/\mathcal J|^2>1$ in this model.

Surprisingly this enhancement factor $|1/\mathcal J|^2$ is of ``oscillation" nature: it reaches the local maximum $|1/\mathcal J|^2_{\rm max} = 1+2\mu V_a/p^2$ when $p_{in}a = n\pi+\pi/2$ and local minimum $|1/\mathcal J|^2_{\rm min}=1$ when $p_{in}a=n\pi$. The energy gaps between the 1st, the 2nd, the 3rd and the 4th minima are:
\begin{equation}
  \frac{3\pi^2}{2\mu a^2},\quad \frac{5\pi^2}{2\mu a^2},\quad \frac{7\pi^2}{2\mu a^2}\,.
\end{equation}
The energy gaps, which are a manifestation of the observed oscillation period, are approximately determined by the interaction range $a$. For the $N\bar{N}$ scattering, we have $\mu = m_N/2$ and $a \approx 1/m_\pi$, and this gives the 1st gap $\Delta E_{1} \approx 0.6$ GeV. This is close to the value observed in experiment.
In addition, the peaks of $|1/\mathcal J|^2$ decrease with the increasing energies. This feature describes the damping behavior of the oscillation observed in the experiment.

\begin{figure}[hbtp]
  \centering
  \hspace{-1.5em}\includegraphics[width=9cm]{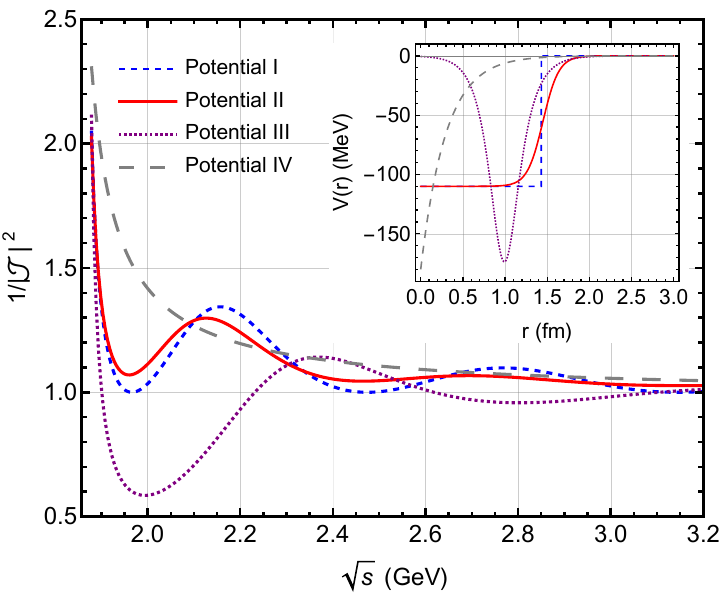}
  \caption{Different potentials and their FSI factors. Potential I: the rectangular well with range $a=1.43$ fm and potential depth $V_0=110$ MeV. Potential II: $V=-V_0/(1+r e^{\frac{r-R}{a}})$ with $V_0=110$ MeV, $R=1.49$ fm and $a=0.1$ fm. Potential III: $V=-(V_0 e^{-r/a_1})/(1+e^{\frac{-r+R}{a_2}})$ with $V_0=140$ GeV, $R=0.99$ fm, $a_1=0.16$ fm and $a_2=0.08$ fm. Potential IV: $V=-V_0 e^{-r/a}$ with $V_0=180$ MeV and $a=0.30$ fm.}\label{fig:factor}
\end{figure}

In Fig.~\ref{fig:factor}, we show the FSI factors $1/|\mathcal{J}|^2$ of several other potentials together with the simple rectangular potential well (Potential I).
The FSI factor of Potential II, a smeared rectangular potential well with smooth edge, also exhibit oscillatory feature but have a weaker effect than Potential I.
Potential III is a smooth potential well located around 1.0 fm and exhibit oscillatory behavior as well.
So it is interesting that a potential of similar interaction range with the rectangular well does result into oscillatory features of form factors.
However, we found that a simple exponential potential (Potential IV in Fig.~\ref{fig:factor}) of the form $V=-V_0 e^{-r/a}$ does not induce any oscillation behavior.
The genuine form of $N\bar{N}$ potential is complicated and unclear, especially for the short range part where annihilation dynamics take place. It is premature and beyond the scope of this work to investigate which kind of potential is more appropriate for the treatment of FSI.
As a compromise, we will adopt the simple framework of rectangular well in the following.

\noindent{\it Description for the effective form factor $G_{\mathrm{eff}}$.}---It is common to express the cross section data in terms of the effective form factor which can be obtained as
\begin{equation}\label{eq:Geff}
  |G_{\mathrm{eff}}(s)| = \sqrt{\frac{3s}{4\pi\alpha^2\beta C(1+2m_N^2/s)} \, \sigma_{e^+e^-\to N\bar{N}} }\,.
\end{equation}
It can be factorized with the $N\bar N$ rescattering correction $1/|\mathcal J|$ and the form factor $G_0$ without FSI:
\begin{equation}
  |G_{\mathrm{eff}}(s)|=\frac{1}{|\mathcal J|}\,G_0(s)\,,
\end{equation}
where $G_0$ describes the short range production and determines the main feature of $G_{\mathrm{eff}}$ in timelike region. Following Ref.~\cite{BESIII:2019hdp}, it can be parametrized as
\begin{equation}\label{eq:G0}
  G_0(s) = \frac{\mathcal{A}}{(1+s/m_a^2)\,[1-s/(0.71~{\rm GeV^2})]^2}\,,
\end{equation}
where $m_a^2 = 7.72$ GeV$^2$ and $\mathcal{A}$ is a constant. With fitted $\mathcal{A}_p=9.37$ and $\mathcal{A}_n = 5.8$, $G_0$ can give excellent overall descriptions of effective form factors for nucleons over a wide range. By subtracting the smooth continuum $G_0$ from $|G_{\mathrm{eff}}|$, we get the oscillation behavior
\begin{equation}
  G^{osc}(s)=|G_{\mathrm{eff}}|-G_0=\left(\frac1{|\mathcal J|}-1\right)G_0(s)\,.
\end{equation}

In the previous section we showed that a simple rectangular potential well already leads to an oscillatory feature, and the enhancement factor is larger than 1 in that case, meaning that $G^{osc}$ will be positive with the attractive interaction. To get the observed oscillatory structure, we need to make the potential a bit more complicated by using
\begin{equation}\label{eq:two_rectangle}
  V(r) = \begin{cases}
        -V_r & 0\leqslant r < a_r \\
       -V_a & a_r \leqslant r < a \\
        0    & r \geqslant a
      \end{cases}
      \,,
\end{equation}
where $0<V_r<V_a$ and we take $a_r=0.5$ fm.

\begin{table}[h]
  \centering
  \caption{Parameters for $p\bar{p}$ and $n\bar{n}$ potentials in Eq.~\eqref{eq:two_rectangle}.}\label{tab:parameter}
  \begin{ruledtabular}
  \begin{tabular}{ccccccc}
    $N\bar{N}$       & $a_r$ (fm) & $V_r$ (MeV) & $a$ (fm) & $V_a$ (MeV)  \\
    \hline
    $p\bar{p}$       & 0.5        & 50          & 1.6      & 90           \\
    $n\bar{n}$       & 0.5        & 400         & 1.4      & 650          \\
  \end{tabular}
  \end{ruledtabular}
\end{table}

\begin{figure*}[htbp]
  \centering
 \hspace{-1.5em} \subfigure{
  \includegraphics[width=8.8cm]{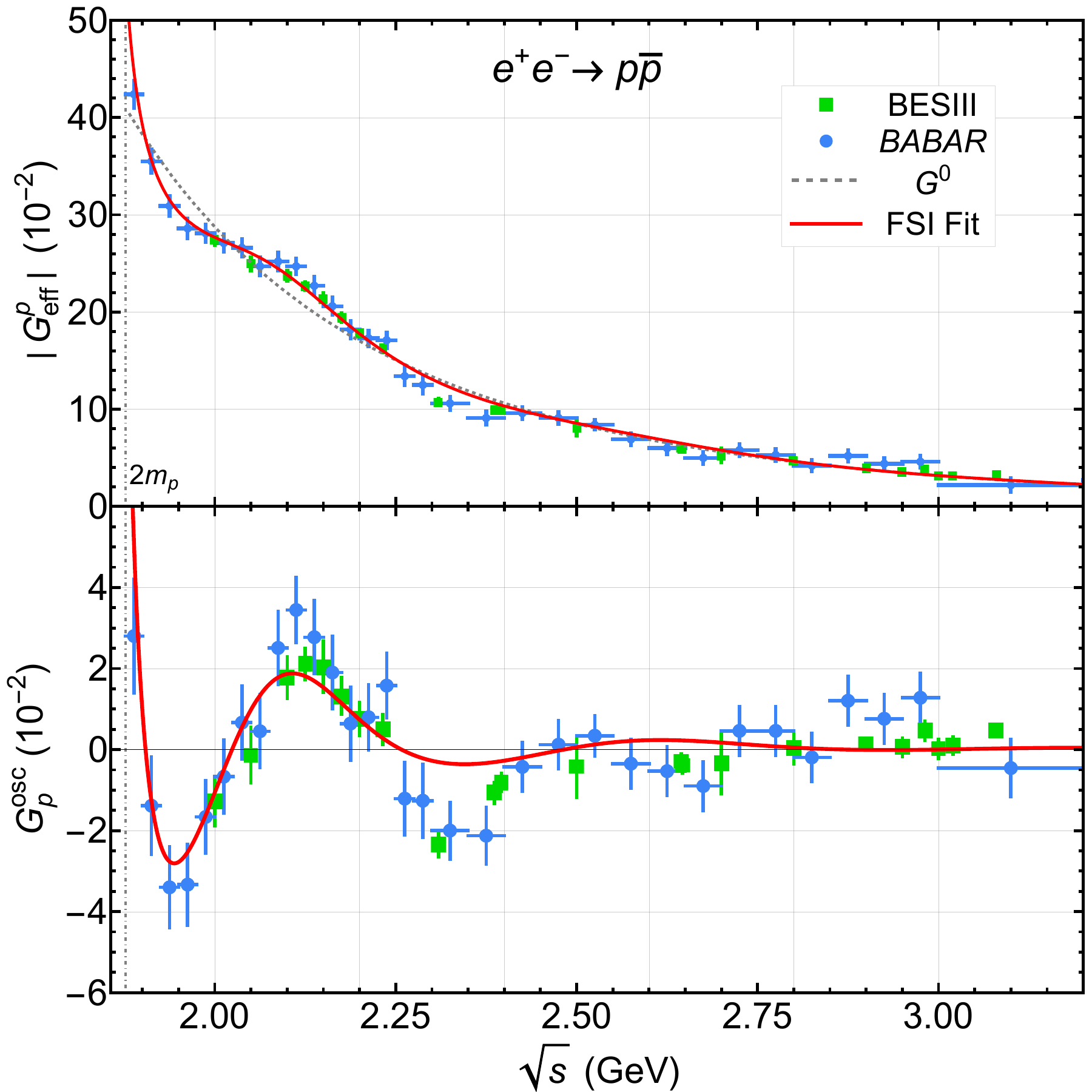}}
  \subfigure{
  \includegraphics[width=8.8cm]{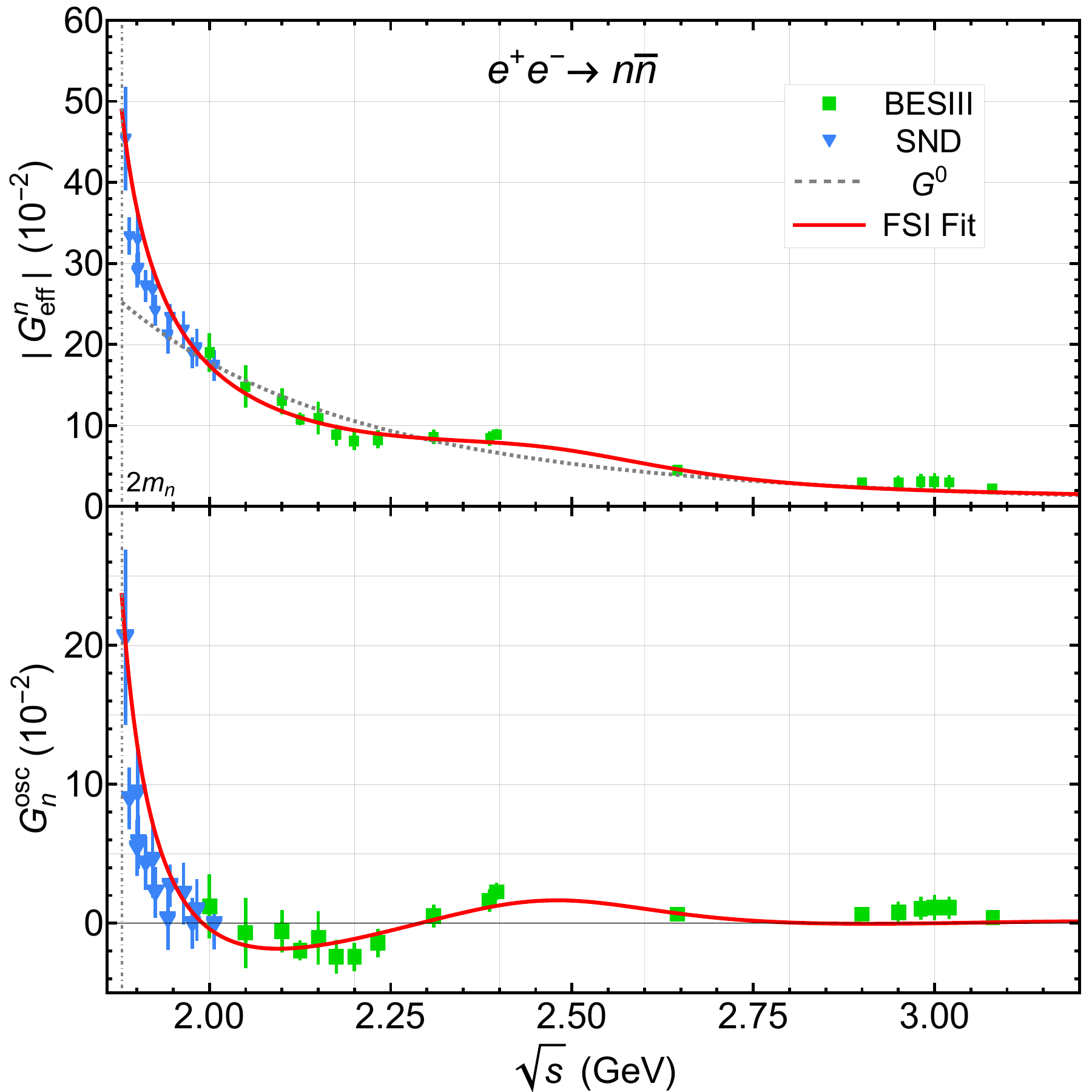}}
  \caption{The effective form factors for protons (left) and neutrons (right) in the timelike region. At the top are the effective form factors $|G_{\rm eff}|$, while at the bottom is the oscillation part $G^{osc}$ after subtracting the smooth background part $G_0$ (gray dashed line). The threshold positions $\sqrt{s}=2m_N$ are marked as gray dot-dashed vertical lines. In the figure, we only include BESIII~\cite{BESIII:2019hdp} (green square) and {\it BABAR}~\cite{Bianconi:2015owa} (blue circle) data for $p\bar{p}$ and  BESIII~\cite{BESIII:2021tbq} (green square) and SND~\cite{SND:2022wdb} (blue triangle) data considering the precision and covering range. }\label{fig:FSI_fit}
\end{figure*}

We show the FSI effect with such potentials on the $p\bar{p}$ and $n\bar{n}$ effective form factors in Fig.~\ref{fig:FSI_fit}. The corresponding parameters are listed in Table.~\ref{tab:parameter}. The overall oscillatory behavior is well reproduced by the FSI effect with the interaction range $a$ about $1.4\sim 1.6$ fm.

The details of the description of the effective form factors depend on the choice of the continuum part $G_0$, which we still do not understand very well, because the formation process involves the complicated hadronization and other difficulties. Perhaps better descriptions can be obtained for $|G_{\rm eff}|$ by using different $G_0$ rather than the same as in the experimental collaborations.

\begin{figure*}[htbp]
  \centering
  \hspace{-1.5em}\includegraphics[width=8.8cm]{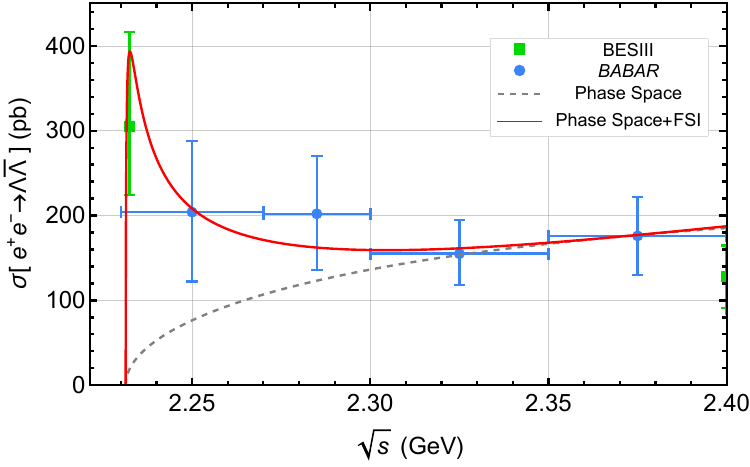}
 \includegraphics[width=8.8cm]{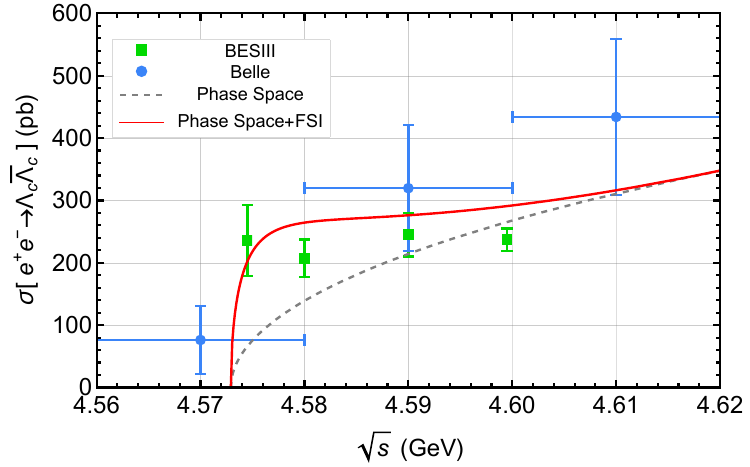}
  \caption{The cross sections near the thresholds for $e^+e^-\to \Lambda\bar\Lambda$~\cite{BESIII:2017hyw,BaBar:2007fsu} and $e^+e^-\to \Lambda_c\bar\Lambda_c$~\cite{BESIII:2017kqg,Belle:2008xmh}. The threshold data deviate far away from the fits (dashed lines) proportional to the phase space distributions. The fits (solid lines) improve after including the FSI enhancement factor given by Eq.~\eqref{eq:enhancement} with $a=1/m_\pi$ and appropriately adjusted $V_0$.
}\label{fig:enhancement}
\end{figure*}
\noindent{\it Threshold enhancement of time-like form factors.}---From Fig.~\ref{fig:FSI_fit}, it can be seen that the neutron effective form factor has a stronger enhancement near the threshold than that for the proton. The SND measurement observed the enhancement on the neutron cross section just above threshold at $\sqrt{s}-2m_n \approx 5$ MeV~\cite{SND:2022wdb}, which contradicts the naive phase space expectation.
The enhancement of nucleon form factors have long been discussed in $N\bar{N}\to e^+e^-$ process and similar enhancement phenomenon in baryon pair production in $e^+e^-$ annihilation should also exist~\cite{Dalkarov:1978as,Dalkarov:1991gw,Dalkarov:1979ps,Dalkarov:1989xs,Dalkarov:2009yf,Shapiro:1978wi,Dmitriev:2007zz,Milstein:2022tfx}.
In the discussion here, this can be seen with a simple rectangular well potential, where the FSI factor near the threshold $1/|\mathcal J|_{p\to 0}\to 1/\mathrm{cos}^2\left(\sqrt{2\mu V_a}a\right)$ has an attractive squared-well potential. With suitable $V_a$ and $a$, $1/|\mathcal J|_{p\to 0}$ can lead to very large enhancement near the threshold.

In addition to the $N\bar{N}$ production, other baryon-antibaryon pair productions near thresholds have also been measured~\cite{BESIII:2017hyw,BESIII:2017kqg,BESIII:2020ktn,BESIII:2021aer,BESIII:2020uqk,BESIII:2021rkn}. Abnormally large cross sections are observed in $e^+e^-\to \Lambda\bar{\Lambda}$ near the threshold $\left(\sqrt{s}-2m_\Lambda\right) \approx 1$ MeV~\cite{BESIII:2017hyw} and possibly $e^+e^-\to \Lambda_c\bar{\Lambda}_c$ at $\left(\sqrt{s}-2m_{\Lambda_c}\right) \approx 1.58$ MeV~\cite{BESIII:2017kqg}. However, no such phenomenon were found in the $\Xi\bar{\Xi}$~\cite{BESIII:2020ktn,BESIII:2021aer} and $\Sigma\bar{\Sigma}$~\cite{BESIII:2020uqk,BESIII:2021rkn,BESIII:2020uqk} productions. These have been studied with different approaches~\cite{Dmitriev:2007zz,Dmitriev:2013xla,Li:2021lvs,Haidenbauer:2014kja,Haidenbauer:2015yka,Cao:2018kos,Yang:2019mzq,Haidenbauer:2020wyp,Dai:2017fwx,Salnikov:2023zyt}. Our framework can provide a naturally unified explanation for the near-threshold enhancements, which are either large or almost nonexistent, based on the different FSIs between these baryon-antibaryon pairs. We can nicely fit the enhanced near-threshold cross sections for productions of $\Lambda\bar\Lambda$ and $\Lambda_c\bar\Lambda_c$ with simple squared-well potentials as shown in Fig. \ref{fig:enhancement}.

\noindent{\it Summary.}---We have introduced a simple framework to deal with the damped oscillation observed in the electromagnetic form factors of nucleons which is measured in $e^+e^-\to N\bar N$. The FSI effect is important for this phenomenon and leads to a factor consisting of the $N\bar N$ Jost function with the distorted-wave Born approximation. Using the squared-well potentials, the FSI factors are naturally damped oscillatory and the experimental data can be well explained. It can be easily extended to interpret the threshold enhancements on the
production cross sections for the $n\bar n$, $\Lambda\bar\Lambda$, and $\Lambda_c\bar\Lambda_c$, and moreover, same approach can also give inapparent enhancements in other channels just with a proper adjustment of the FSI parameters.

The oscillation phenomenon could also exist in the production of hyperon-antihyperon and other pairs, which is an interesting topic in both experiment and theory~\cite{Dai:2021yqr}. With the better understanding of the form factors in the future, we hope that it will also be possible to reveal the related structures of nucleons and other baryons.

\begin{acknowledgments}
Z.-W. Liu thanks Prof. Anthony W. Thomas, Derek B. Leinweber, and Lei Chang for useful discussions. This work is supported by the China National Funds for Distinguished Young Scientists under Grant No. 11825503, the National Key Research and Development Program of China under Contract No. 2020YFA0406400, the 111 Project under Grant No. B20063, the National Natural Science Foundation of China under Grant Nos. 12175091, 11965016, 12047501, 12247101 and U2032109, and the Project for Top-notch Innovative Talents of Gansu Province.
\end{acknowledgments}
\bibliography{oscillation.R1}

\clearpage
\section*{\large Supplemental materials}
\label{supp_mat}

\setcounter{equation}{0}
\setcounter{figure}{0}
\setcounter{table}{0}
\renewcommand{\theequation}{S\arabic{equation}}
\renewcommand{\thefigure}{S\arabic{figure}}
\renewcommand{\thetable}{S\arabic{table}}

we notice that the BESIII collaboration reported their new measurement on the cross section of $e^+e^-\to\Lambda\bar{\Lambda}$ via ISR very recently~\cite{BESIII:2023ioy}. As shown in Fig.~\ref{fig:enhancement_La}, the new data are consistent with our earlier description.
\begin{figure}[hbt]
  \centering
  \includegraphics[width=8cm]{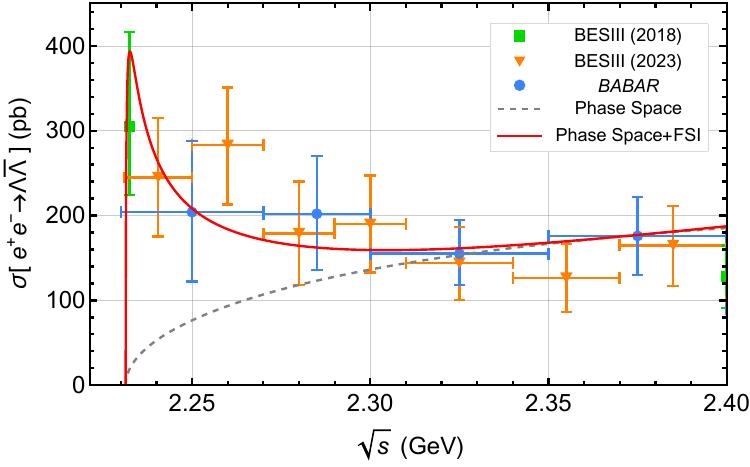}
  \caption{The cross sections near the thresholds for $e^+e^-\to \Lambda\bar\Lambda$~\cite{BESIII:2017hyw,BaBar:2007fsu,BESIII:2023ioy}. The curves and captions are same as the left figure of Fig. \ref{fig:enhancement} but with new yellow data from Ref. \cite{BESIII:2023ioy} added.
}\label{fig:enhancement_La}
\end{figure} 
\clearpage
\end{document}